\documentclass[final,5p,times,twocolumn]{elsarticle}
\usepackage{color}
\usepackage{amsmath}
\usepackage{amssymb}
\usepackage{subfig}
\usepackage{graphicx}
\usepackage[colorlinks=true,linkcolor=black, citecolor=blue, urlcolor=blue]{hyperref}
\usepackage{lineno,hyperref}
\biboptions{numbers,sort&compress}   
\modulolinenumbers[5]

\begin{document}

\begin{frontmatter}

\title{Single array of magnetic vortex disks uses in-plane anisotropy to create different logic gates}

\author{H. Vigo-Cotrina}
\address{Centro Brasileiro de Pesquisas F\'{\i}sicas, 22290-180,  Rio de Janeiro, RJ, Brazil}

\author{A.P. Guimar\~aes}
\address{Centro Brasileiro de Pesquisas F\'{\i}sicas, 22290-180,  Rio de Janeiro, RJ, Brazil}




\begin{abstract}
Using micromagnetic simulation, we show that in-plane uniaxial magnetic anisotropy (IPUA) can be used to obtain FAN-OUT, AND and OR gates in an array of coupled disks with magnetic vortex configuration. First, we studied the influence of the direction of application of the IPUA on the energy transfer time ($\tau$) between two identical coupled nanodisks. We found that when the direction of the IPUA is along  the x axis the magnetic interaction increases, allowing shorter values of $\tau$,  while the IPUA along the y direction has the opposite effect. The magnetic interactions between the nanodisks along x and y directions (the coupling integrals) as a function of the uniaxial anisotropy constant (K$_{\sigma}$) were obtained using a simple dipolar model. Next, we demonstrated that choosing a suitable direction of application of the IPUA, it is possible to create several different logic gates with a single array of coupled nanodisks.
\end{abstract}


\end{frontmatter}


\section{Introduction}\label{introduction}

The magnetic vortex configuration is a ground state characterized by a curling magnetization in the plane,  and a small region (core vortex) in the center, where of magnetization is out the plane \cite{Guslienko:2008, Guimaraes:2009}. From these characteristics,  two properties are defined: the circulation C and the polarity p. The circulation is C = +1 when the curling direction is counterclockwise (CCW) and C = -1 when it is clockwise (CW). The polarity is p = +1 when the core points in the +z direction and p = -1 in the -z direction. The magnetic vortex dynamics is characterized by an eigenfrequency (gyrotropic frequency) in the sub-gigahertz range \cite{Guslienko:2008, Guimaraes:2009, Guslienko:2002, Guslienko:2006}. This frequency depends on the intrinsic parameters of the material and on the ratio $\beta$ of the thickness to the radius of the disk ($\beta$ = L/R) \cite{Guslienko:2008}.\\
\indent The use of the magnetic vortices in logic gates is one of the potential technological applications in spintronics \cite{Jung2012,Bowden2010,Barman2011,Omari:2014}. Controlling parameters such as the polarity and circulation, different logical configurations can be obtained, as already demonstrated by Jung \textit{et al.} \cite{Jung2012}, and controlling the magnetic interaction through the separation distance between the disks, it is possible to perform fan-out operations \cite{Barman2011}. Therefore, the control of these parameters is of vital importance for the construction of logic gates. Both p as C can be controlled with rotating magnetic fields \cite{Fior:2016,Jung2012}, and the magnetic interaction can be modified without the need to change the separation distance between the disks.\\
\indent One of the ways of obtaining this goal is by the application of  perpendicular magnetic fields (PMF) to the plane of the disks \cite{Yoo:2011}. This method may not be very effective, since the PMF distorts the  core profile of the vortex, which is where the information bits are usually stored; therefore, the search for other mechanisms is an open topic. In this sense, the influence of the in-plane uniaxial anisotropy (IPUA) on the dynamics of magnetic vortices is gaining interest in recent years \cite{Roy:2013,Helmunt2016}. Experimentally, the IPUA can be induced by voltage-induced strain via a piezoelectric transducer (PZT) \cite{Ostler:2015, Helmunt2016}. The advantage of using the IPUA, is that this does not alter the core profile of the vortex, making it a practical tool to control the dynamics of the magnetic vortex \cite{Roy:2013, Helmunt2016}.\\
\indent When disks with magnetic vortex configuration are coupled, energy is transfered periodically between them \cite{Jung:2011}. The energy transfer is characterized by an energy transfer time parameter $\tau$. Therefore, controlling this parameter is very important to affect this transfer. In a previous work, we have shown that applying the IPUA along the x axis allows obtaining shorter energy transfer times \cite{Helmunt2016}.\\
\indent The goal of this work is to study the influence of changing the direction of application of the IPUA in a pair of identical nanodisks, and using the IPUA in order to obtain different logical configurations with an array of coupled nanodisks. We used micromagnetic simulations and a simple analytical dipolar model in order to obtain the interactions along the x and y directions (the coupling integrals). We used the open source software Mumax \cite{Vansteenkiste:2014}, with cell size of 2 $\times$ 2 $\times$ 7$\:$nm$^3$; the magnetostrictive material used was Galfenol (FeGa) with parameters \cite{SungHwan:2010, Summers:2007}: saturation magnetization M$_s$ = 1.360 $\times$ 10$^6$ A/m$^2$, exchange stiffness \textit{A} = 14 $\times$ 10$^{-12}$ $\:$J/m  and a typical damping constant $\alpha$ = 0.01. The anisotropy constant (K$_{\sigma}$) varied from 0 to 58500 J/m$^3$. The magnetoelastic energy was included in the micromagnetic simulation as an uniaxial anisotropy energy \cite{Roy:2013,Helmunt2016}.

\section{Results and discussion}
\subsection{Two coupled disks}\label{twocoupled}
\subsubsection{Micromagnetic simulation}
\begin{figure}[h!]
\centering
\includegraphics[width=1\columnwidth]{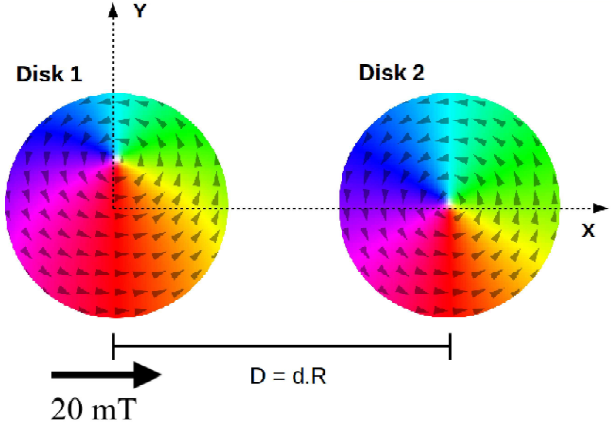}
\caption{Coupled nanodisks with magnetic vortex configuration separated by a center-to-center distance D;  d = D/R is the reduced distance. The vortex core of disk 1 was displaced by an in plane static magnetic field $\mu_0$H = 20$\,$mT in the +x direction. This is the typical initial configuration used to perform the study of the energy transfer time between two disks.}\label{2discos}
\end{figure}

\begin{figure}[h!]
\centering
\includegraphics[width=1\columnwidth]{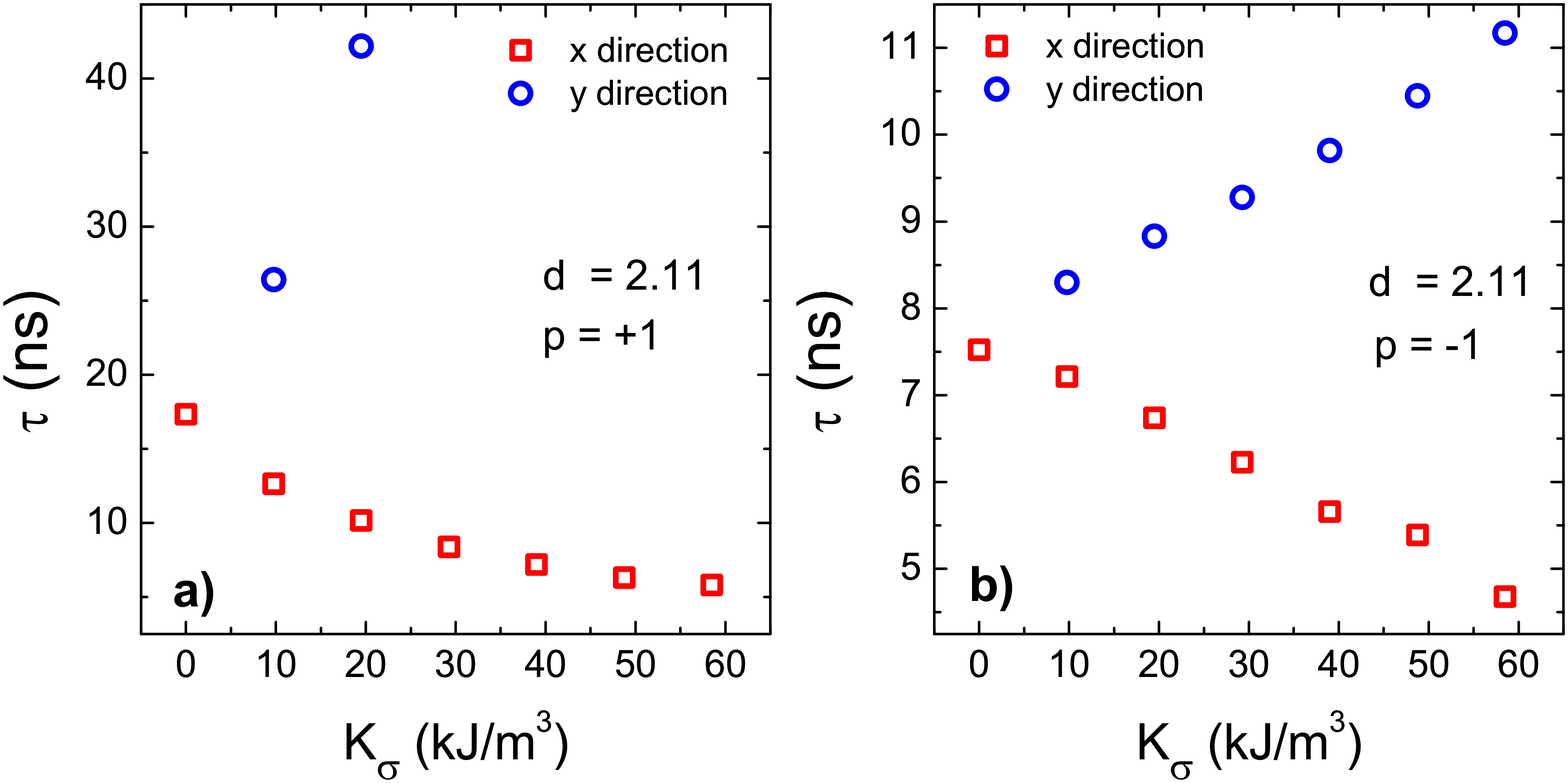}
\includegraphics[width=1\columnwidth]{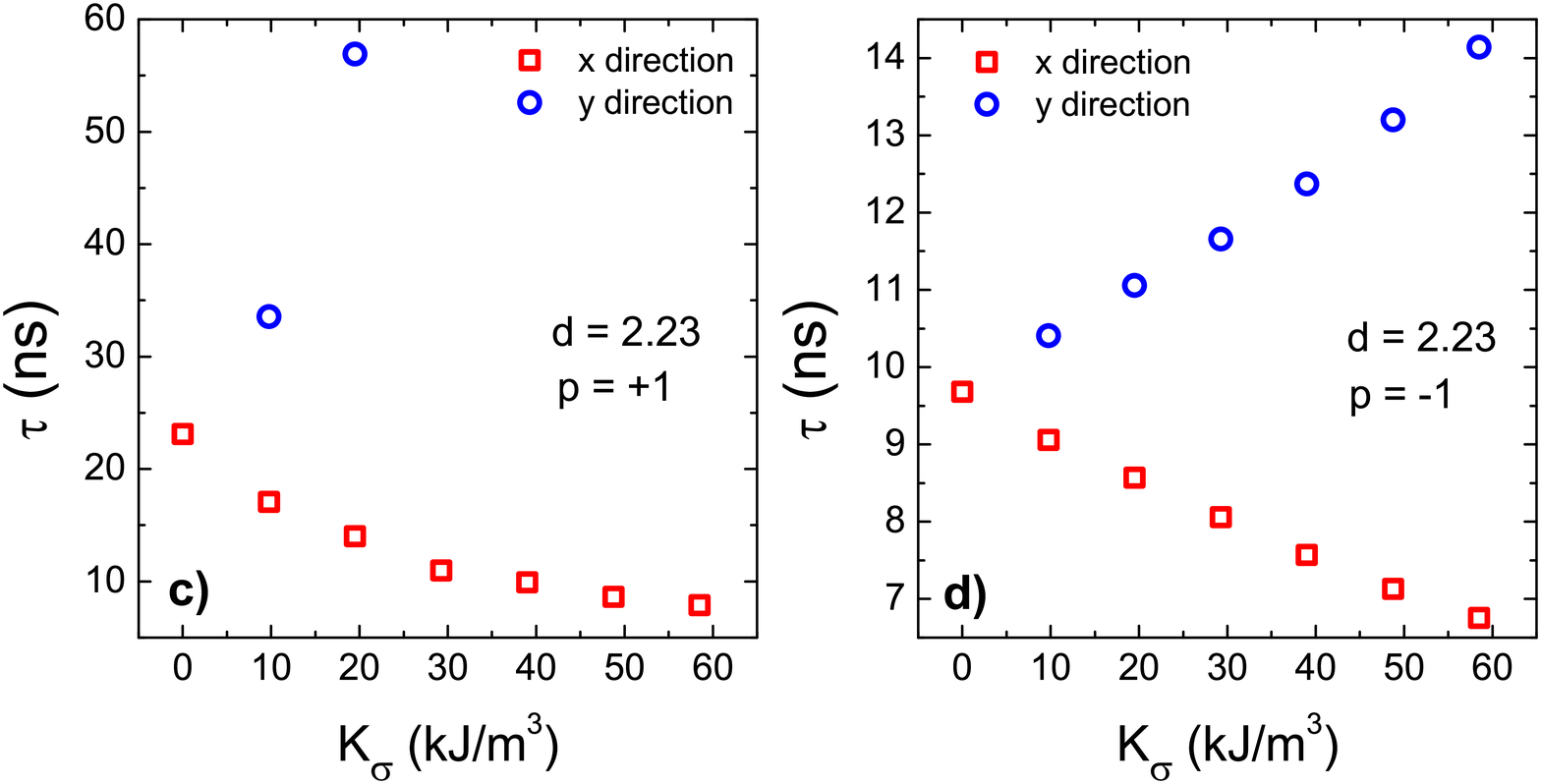}
\caption{Variation of $\tau$ with respect to K$_{\sigma}$ for a reduced distance d = D/R = 2.11 (a-b) and d = D/R = 2.33 (c-d) for the cases p = +1 and p = -1}\label{tau}
\end{figure}

\indent We considered a system of two coupled identical disks, located along the x-axis, with thickness L = 7$\:$nm, diameters 256$\:$nm, and separated by a center to center distance D (Fig. \ref{2discos}). In order to induce gyrotropic motion we have applied an in-plane static magnetic field in the x direction for a few nanoseconds, only on disk 1 to displace the vortex core from the equilibrium position (center of the disk), and using a large damping $\alpha = 1$ for faster convergence. Then this field was removed and the IPUA was applied to both disks, a typical value of $\alpha = 0.01$ was used, allowing the vortex core to perform the gyrotropic motion. The movement of the vortex core in the disk 2 is induced by the transfer of energy from disk 1. This process is repeated for each value of K$_{\sigma}$.\\
\indent Unlike our previous work, we have also considered the case when the uniaxial anisotropy is in the direction of the y axis. The values of $\tau$  depending on the K$_{\sigma}$ are shown in Fig. \ref{tau}. This dependence of $\tau$ with K$_{\sigma}$ is the opposite of that obtained in our previous work, where the direction of the IPUA was applied along the x-axis \cite{Helmunt2016}.

When the direction of application of the IPUA is along the y axis (IPUA-y), the values of $\tau$ increase, while when the direction is along the x axis (IPUA-x), $\tau$ decreases. For example, considering IPUA-x, for the case p = p$_1$.p$_2$ = +1 and a reduced distance d = D/R = 2.11, $\tau$ is reduced by approximately 66$\%$ from $\tau$ = 17.3$\,$ns (K$_{\sigma}$ = 0$\,$kJ/m$^3$) to $\tau$ = 5.8$\,$ns (K$_{\sigma}$ = 58.5$\,$kJ/m$^3$) and for the case p = -1, $\tau$ is reduced approximately by 40$\%$ from $\tau$ = 7.5$\,$ns (K$_{\sigma}$ = 0$\,$kJ/m$^3$) to $\tau$ = 4.6$\,$ns (K$_{\sigma}$ = 58.5$\,$kJ/m$^3$). This reduction in the $\tau$ values is  appreciable and show the efficiency of using IPUA as a tool for the control of the energy transfer time, as we have already demonstrated in our previous work \cite{Helmunt2016}. Now, for IPUA-y and a reduced distance d = 2.11, $\tau$ increases approximately 140$\%$, from $\tau$ = 17.3$\,$ns (K$_{\sigma}$ = 0$\,$kJ/m$^3$) to $\tau$ = 42$\,$ns (K$_{\sigma}$ = 19.5$\,$kJ/m$^3$) for the case p = +1, and an increase of approximately 48$\%$, from $\tau$ = 7.5$\,$ns (K$_{\sigma}$ = 0$\,$kJ/m$^3$) to $\tau$ = 11.2$\,$ns (K$_{\sigma}$ = 58.5$\,$kJ/m$^3$), for the p = -1 case. In every case, the values of $\tau$ were obtained from the temporal dependence of the energy density of each disk, considering that $\tau$ is the time in which the energy of disk 1 reaches the minimum value for the first time \cite{Jung:2011}.

\indent As can be seen in Fig. \ref{tau}, we have considered only three values of K$_{\sigma}$ (0, 9.75 and 19.5$\,$kJ/m$^3$) for IPUA-y, d = D/R = 2.11, and p = +1,  For larger values of K$_{\sigma}$, a simple free induction decay (FID), with a decay time of approximately 100$\,$ns and the same gyrotropic frequency (thus there is no frequency splitting), is observed in both disks. This means that the energy of disk 1 is fully transferred to disk 2 for a value of $\tau$ comparable to the decay time of the FID.  During this stage, the vortex core of disk 2 performs a gyrotropic motion of negligible amplitude.\\
\indent The initial amplitude of the FID in disk 1 is due to the initial displacement of the vortex core by the application of the in-plane magnetic field. During this stage, although no  magnetic field is applied in disk 2, the vortex core is also displaced due to the magnetic interaction between the disks \cite{Suk2011}. This interaction is due to the presence of surface and/or volume charges in disk 1 resulting from the displacement of the vortex core by the application of the in-plane magnetic field \cite{Sinnecker:2014, Suk2011}. The displacement in disk 2 is  negligible in comparison to the displacement in disk 1, and is of the order of 15$\,$nm. 
\indent For IPUA-y and p = -1, we have used 58.5$\,$kJ/m$^3$ as the maximum value of K$_\sigma$, as already mentioned in section \ref{introduction}. For larger values of K$_{\sigma}$, the magnetic vortex configuration is not stable.\\
\indent The reason why it is still observed an energy transfer, even using our maximum allowed value of K$_{\sigma}$, is because the magnetic interaction between the disks is stronger in comparison with the case p = +1, therefore,  high values of K$_{\sigma}$ are required to decrease the magnetic interaction between the disks.

\subsubsection{Analytical dipolar model}
In order to clarify the physical meaning of the influence of the IPUA on the magnetic interactions, we used a simple dipolar model, also used in  our previous work \cite{Helmunt2016}. The  magnetic interaction energy ($W_{int}$) between two disks with magnetic vortex configuration is given by \cite{Sukhostavets:2011}:

\begin{equation}\label{interacciondipolar1}
W_{int} = C_1C_2(\eta_xx_1x_2 + \eta_yy_1y_2),
\end{equation}
where $\eta_x$ and $\eta_y$ are the interactions between the disks along x and y directions (the coupling integrals), C$_1$ and C$_2$ are the circulations in disk 1 and disk 2, and $\bf{X}_i = (x_i,y_i)$, with i = 1,2 is the vector position of the vortex core of the disks.\\ 
\indent The coupling integrals can be expressed as a sum of contributions of high order magnetic interactions, such as  dipole-octupole, octupole-octupole, dipole-triacontadipole, etc. \cite{Sukhostavets:2013}. However, in our model we only consider purely dipole interaction since this is sufficient to explain the influence of IPUA on the magnetic interactions, as we demonstrated in our previous work \cite{Helmunt2016}. Thus, within our dipolar model, the coupling integrals are defined as follows:

\begin{equation}\label{etas}
\eta_x = \eta^* \hspace{1cm}
\eta_y = -2\eta^*,
\end{equation}

\noindent or in their dimensionless forms:

\begin{equation}\label{integraisdeacoplamento}
I_{x,y} = \frac{8\pi}{\mu_0RM_s^2}\eta_{x,y}, 
\end{equation}

\noindent where $\eta^* = \mu_0\lambda^2M_s^2L^2R^2/4\pi D^3$.\\
\indent The $\lambda$ parameter, which takes into account the influence of K$_{\sigma}$, can be obtained from a simultaneous fit of the coupling frequencies f$_{1,2}$ = $\omega_{1,2}/2\pi$ (or eigenfrequencies) as has already been demonstrated by Asmat \textit{et al.} \cite{Asmat:2015}. In order to obtain this parameter, we used the Thiele's equation \cite{Thiele:1973}. Considering a negligible damping, this equation can be written as:

\begin{equation}\label{Thiele}
\textbf{G} \times \frac{d\textbf{X}}{dt} - \frac{\partial W(\textbf{X})}{\partial\textbf{X}} = 0
\end{equation}

\noindent where $\textbf{G}$ is the gyrovector $\textbf{G}$ = -Gp$\hat{z}$, G = 2$\pi \mu_0$LM$_s$/$\gamma$ is the gyrotropic constant, and $\gamma$ = 2.21$\times$10$^5$ $\,$m/As is the gyromagnetic ratio; W($\textbf{X}$) = W(0) + $\frac{1}{2}\kappa\textbf{X}^2$ is the potential energy and $\kappa = 40\pi$M$_s^2$L$^2$/9R is the stiffness coefficient calculated within the side-charge-free model at L/R $<<$ 1 \cite{Guslienko:2008}.\\
\indent In our previously work \cite{Helmunt2016}, from the first variation of the Lagrangian expression for a pair of coupled disks based on the constant of the Thiele's equation, we obtained analytical expression for the coupling frequencies (Eq. (11) in reference \cite{Helmunt2016}), written as: 

\begin{equation}\label{eigenfrequencies}
\omega_{1,2}^p = \sqrt{\omega_0^2 - 2p(\frac{\eta^*}{G})^2 \pm \frac{\eta^*}{G}\omega_0\sqrt{5 - 4p}},
\end{equation} 

\noindent with $\omega_0 = 2\pi f_0$, where f$_0$  is the gyrotropic frequency for an isolated disk.\\
\indent Replacing the expressions for $\textbf{G}$, $\kappa$, f$_0$ and $\eta^*$ in Eq. \ref{eigenfrequencies}, we have:

\begin{equation}\label{iguales}
f_{1,2}^{p = +1} = f_0\sqrt{1 - 0.0103\frac{\lambda^4}{d^6} \pm 0.0718\frac{\lambda^2}{d^3}}
\end{equation}
\begin{equation}\label{diferentes}
f_{1,2}^{p = -1} = f_0\sqrt{1 - 0.0103\frac{\lambda^4}{d^6} \pm 0.2154\frac{\lambda^2}{d^3}}
\end{equation}

\begin{figure}[h!]
\centering
\includegraphics[width=1\columnwidth]{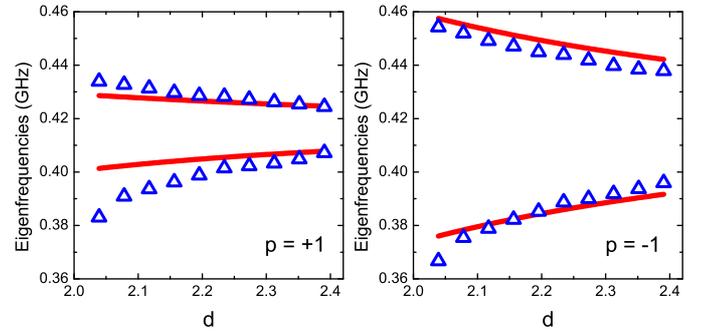}
\caption{Variation of eigenfrequencies f$_{1,2}$ with the reduced distance d = D/R between two disks, for the cases p = p$_1$.p$_2$ = +1 and p = p$_1$.p$_2$ = -1, and considering K$_\sigma$ = 9.75$\,$kJ/m$^3$ (IPUA-x). The blue triangles represent the values obtained from micromagnetic simulations, and the red solid line represents the values obtained from the simultaneous fit using Eq. (\ref{iguales}) and Eq. (\ref{diferentes}).}\label{eigenfre}
\end{figure}

Eq. (\ref{iguales}) and Eq. (\ref{diferentes}) are the analytical expressions for the coupling frequencies (f$_1$ and f$_2$ are the high (+) and low (-) frequencies) within the dipolar approximation  for p = p$_1$.p$_2$ = +1 and p = p$_1$.p$_2$ = -1, respectively. The dependence of the eigenfrequencies on reduced distance d = D/R is shown in Fig. \ref{eigenfre} for  K$_\sigma$ = 9.75$\,$kJ/m$^3$ and IPUA-x.

\begin{figure}[h]
\centering
\includegraphics[width=1\columnwidth]{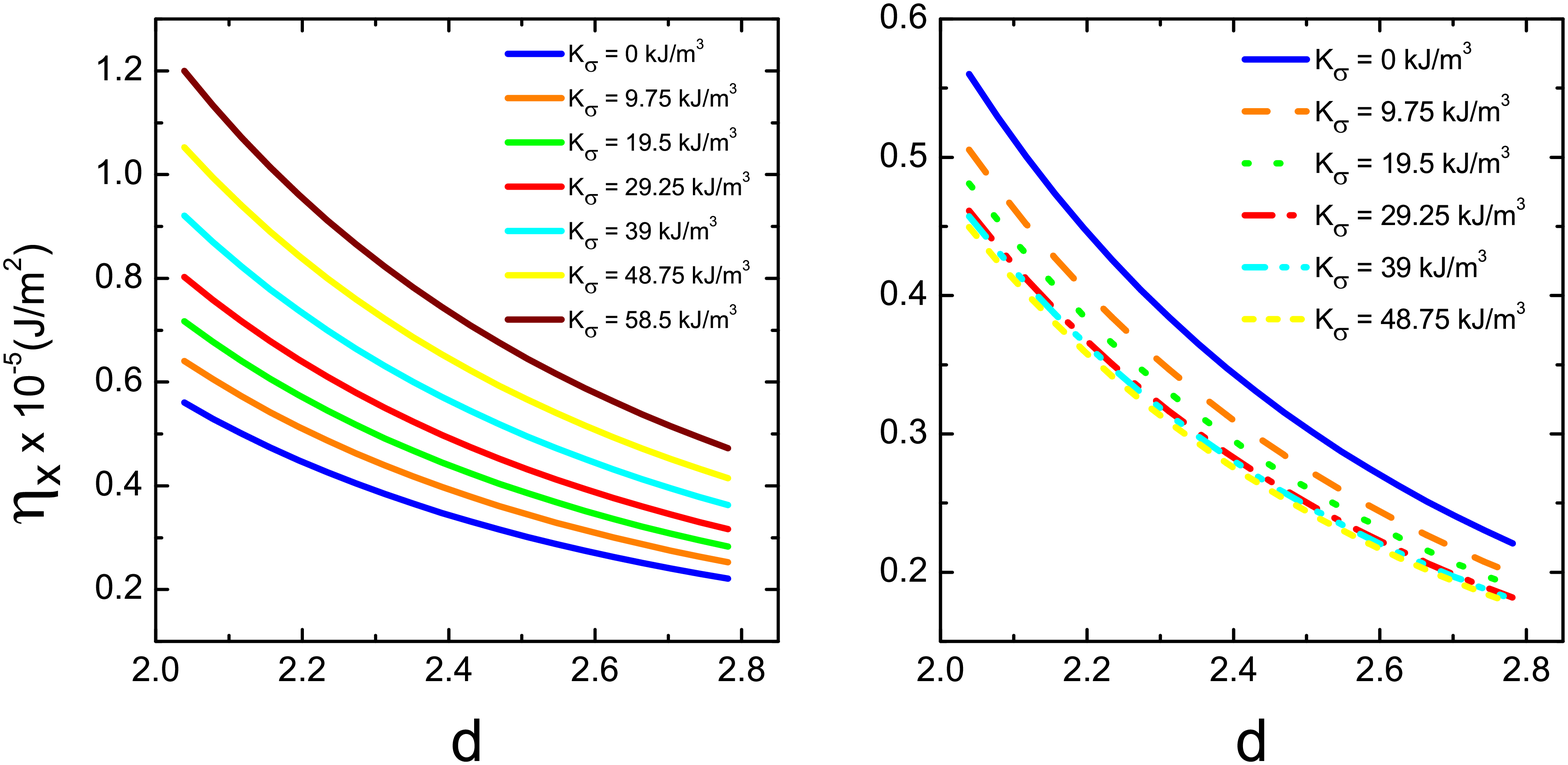}
\caption{Coupling integral $\eta_x$ as a function of the reduced distance d = D/R for disks with L = 7$\,$nm and diameter 256$\,$nm. These results were obtained from Eq. (\ref{etas}) and Eq. (\ref{integraisdeacoplamento}). The solid lines represent the values for the case when the direction of the IPUA is along the x axis (left side), and the dashed and dotted lines represent the values for the cases when the direction of the IPUA is along the y  axis (right side).}\label{Interacciones_x_y}. 
\end{figure}

\indent The values of lambda were obtained from the simultaneous fits to the results shown in Fig. \ref{eigenfre}.\\
\indent A value of $\lambda = 2.77$ is obtained for  K$_\sigma$ = 9.75$\,$kJ/m$^3$. The same procedure is done for each value of K$_\sigma$. We obtained an increases of the $\lambda$ parameter, from $\lambda$ = 2.59 (K$_\sigma$ = 0 $\,$kJ/m$^3$) to $\lambda$ = 3.79 (K$_\sigma$ = 58.5$\,$kJ/m$^3$) for IPUA-x. For IPUA-y \footnote{In this case, the simultaneous fit is not possible, since for p = +1, there is no   frequency splitting for some values of K$_\sigma$ (subsection \ref{twocoupled}). Thus, we have used only the eigenfrequencies obtained for the case p = -1.} we obtained a decrease of the $\lambda$ parameter, from $\lambda$ = 2.59 (K$_\sigma$ = 0 $\,$kJ/m$^3$) to $\lambda$ = 2.32 (K$_\sigma$ = 58.5$\,$kJ/m$^3$). There are discrepancies between the values obtained from micromagnetic simulation and the fit using Eq. (\ref{iguales}). This is expected, since our model considers only dipolar interaction, without consider high-order magnetic interactions as dipole-octupole, octupole-octupole, dipole-triacontadipole, which are important for small separation distances \cite{Sukhostavets:2013, Helmunt2016}. However, the fit is well-behaved when the separation distance between the disks is larger \cite{Helmunt2016}. \\
\indent The dependence of the coupling integral $\eta_x$ on K$_{\sigma}$ and the reduced separation distance d is shown in Fig. \ref{Interacciones_x_y}. For IPUA-x, the coupling integrals increase with the increase of  K$_{\sigma}$, while for IPUA-y, the coupling integrals decrease with the increase of K$_{\sigma}$. The dependence of the coupling integrals on K$_{\sigma}$ and on the reduced separation distance d is shown in Fig. \ref{Interacciones_x_y}. In order to explain the dependence of $\tau$ on K$_{\sigma}$, we used the analytical expression of $\tau$ \cite{Jung:2011}: 

\begin{equation}\label{tau_analitico}
\tau \sim 1/\|\eta_x + p\eta_y\|.
\end{equation}

\noindent Substituting Eq. (\ref{etas}) in Eq. (\ref{tau_analitico}), we have: 

\begin{equation}\label{nuevotau}
\tau \sim 1/\eta^{*} \|1 - 2p\|
\end{equation}

From Eq. (\ref{nuevotau}), we have that the value of $\tau$ is inversely proportional to $\eta^{*}$. Therefore, we have that for IPUA-x, the coupling integrals increase with K$_{\sigma}$ (Eq. (\ref{etas})), thus $\tau$ decreases, and for IPUA-y, the coupling integrals decrease with K$_{\sigma}$, thus $\tau$ increases. This analytical approach applies to both p = +1 and p = -1, and is consistent with the results shown in Fig. (\ref{tau}).\\
\indent Next, we used the influence of the IPUA in order to obtain several logic gates using an array of coupled nanodisks with magnetic vortex configuration.

\subsection{Logic gate applications}\label{gatelogic}
\subsubsection{Fan-out gate}
\begin{figure}[h]
\centering
\includegraphics[width=1\columnwidth]{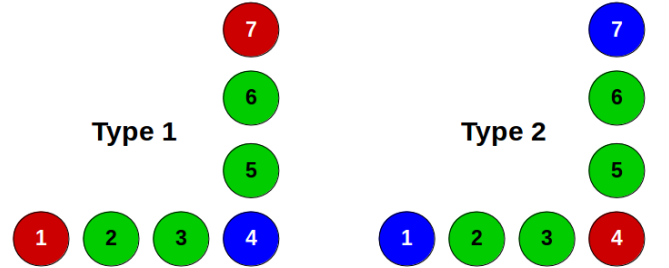}
\caption{Arrays of coupled nanodisks with magnetic vortex configuration. Type 1 is a FAN-OUT gate and Type 2 is an AND/OR gate. Blue color indicates the inputs and red color the outputs.}\label{fanout}
\end{figure}

\begin{figure}[h!]
\centering
\includegraphics[width=1\columnwidth]{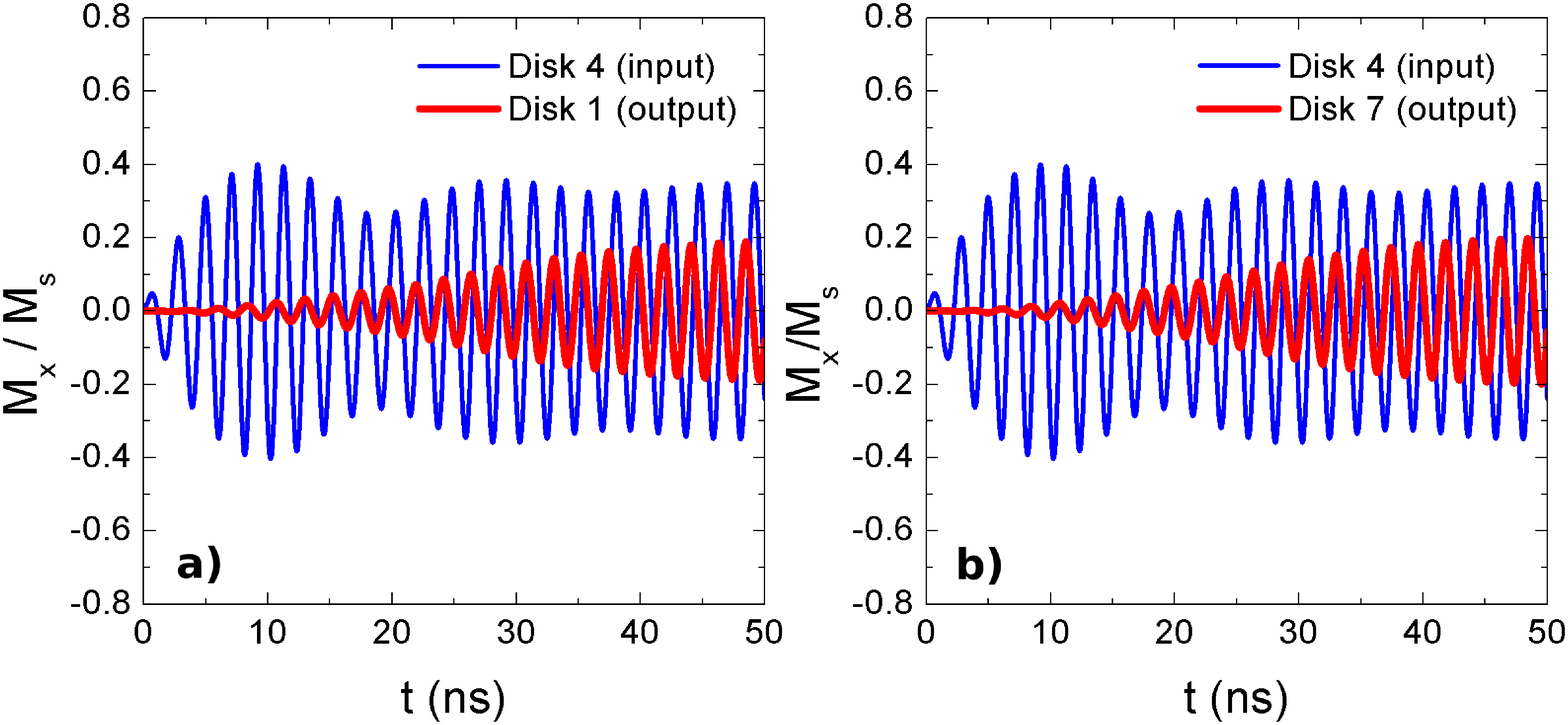}
\includegraphics[width=1\columnwidth]{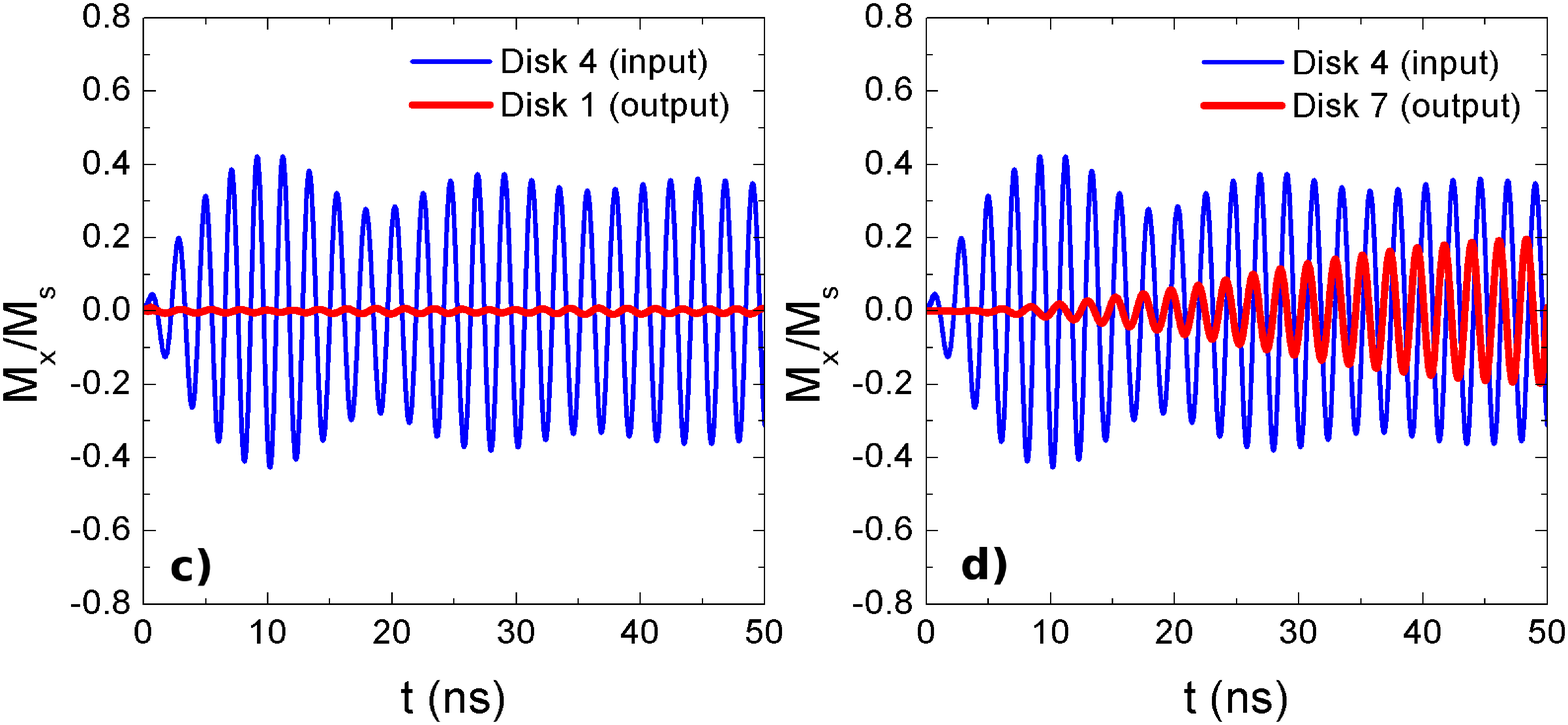}
\includegraphics[width=1\columnwidth]{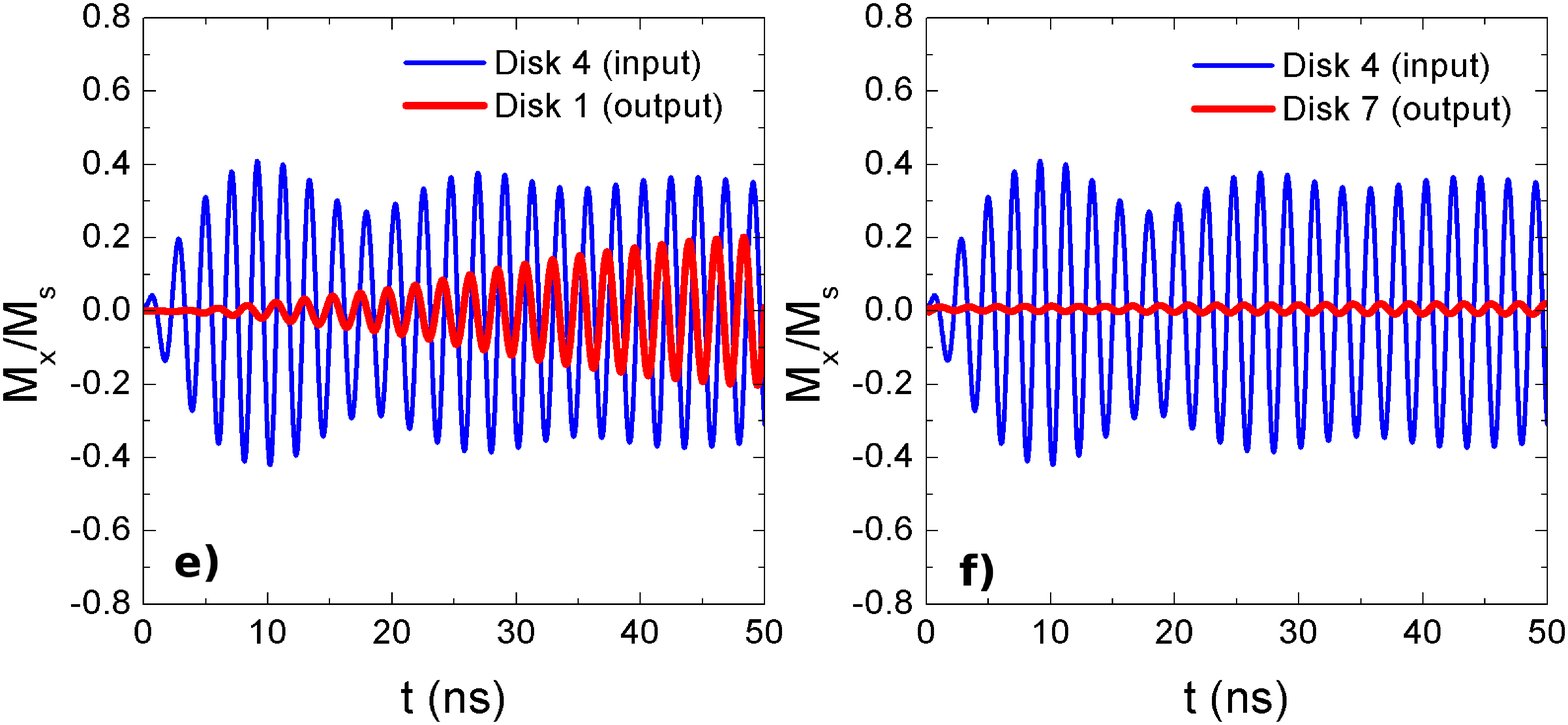}
\includegraphics[width=1\columnwidth]{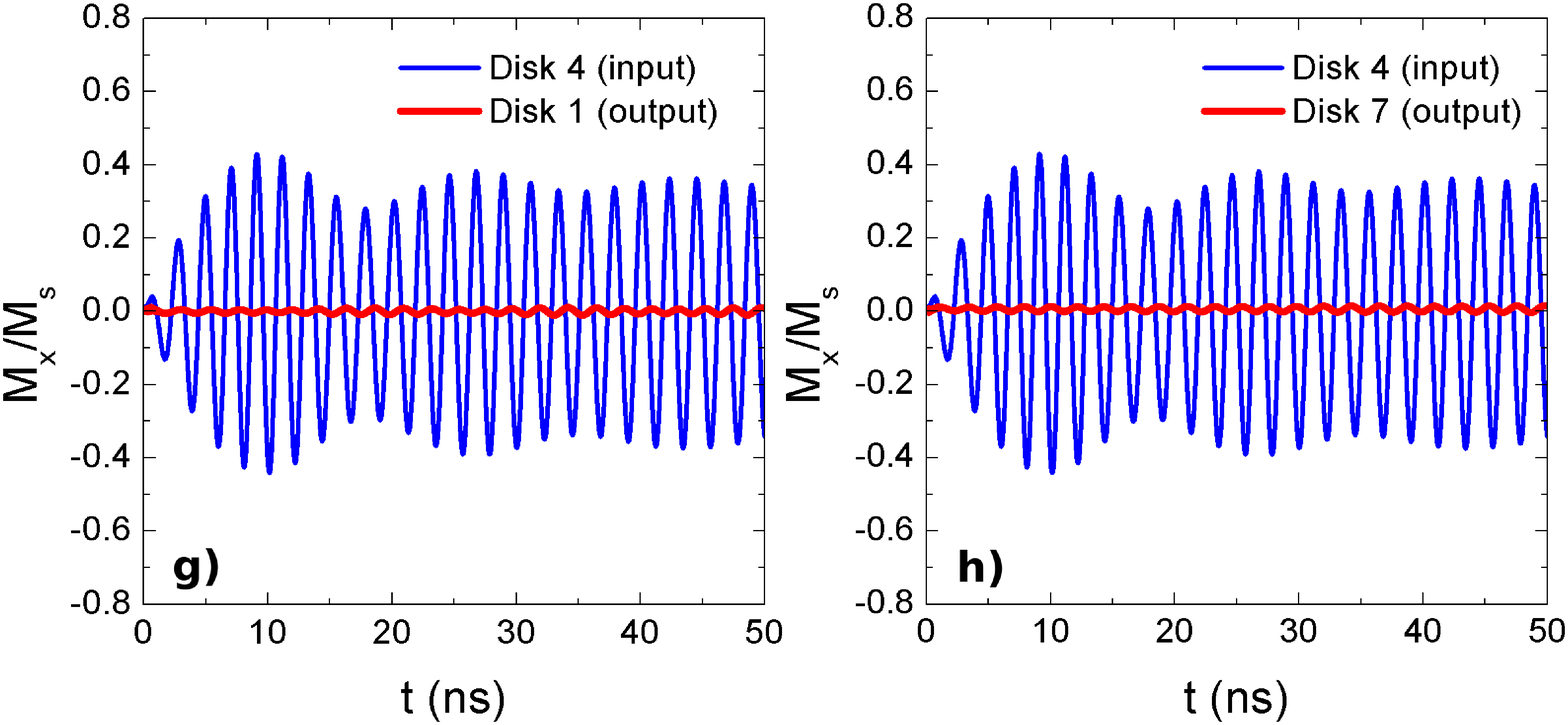}
\caption{Micromagnetic simulation showing a comparison between the intensity levels of the input signal (disk 4) and output signals (disk 1 and disk 7) for a FAN-OUT gate. The signals are M$_x$/M$_s$ as a function of time.}\label{salidas1}
\end{figure}

We used an array of seven identical  disks, as shown in Fig. \ref{fanout} (Type 1), with diameter 240$\,$nm, thickness L = 7$\,$nm and separation distance D = 260$\,$nm between the disks \footnote{The disks can be arranged in different ways to achieve the desired gate. For example, in  ref \cite{Barman2011}  it is shown another type of arrangement to achieve a fan-out gate.}. All disks have positive polarity p = +1 and positive circulation C = +1. The reason for choosing parallel polarities is that the interaction between the disks can be decreased using IPUA, whereas this would be more difficult in the case of having antiparallel polarities (subsection \ref{twocoupled}). We have chosen an L-Shaped (LS) array geometry, because by coupling several LS  it is possible to form a two dimensional network array consisting of nodes, where each node can serve as a nonvolatile memory bit, as has already been proposed by Zhang \textit{et al.} \cite{Zhang:2015}.\\
\indent We considered the disk 4 (blue disk) as an input and disk 1 and disk 7 as outputs (red disks). All the nanodisks have positive polarity and counterclockwise circulation.\\
\indent The type of array  where there is one input and two outputs is known as a fan-out gate \cite{Omari:2014,Zhang:2015,Barman2011}. Since our objective is to observe the output signals on disk 1 and disk 7, the IPUA is applied only to disks 2, 3, 5 and 6. We used in all cases a value K$_\sigma$ = 29.25$\,$kJ/m$^3$. The input signal on  disk 4 is implemented by applying a counterclockwise rotating magnetic field \textbf{B}(t) = B$_0$ $\cos(2\pi f_0t)$ $\hat{x}$ + B$_0$ $\sin(2\pi f_0t)$ $\hat{y}$, where B$_0$ = 2$\,$mT is a magnetic field amplitude and $f_0$ = 0.45$\,$GHz is the gyrotropic frequency of one isolated disk. The same effect could be obtained with a time-dependent polarized current. The value of B$_0$ was chosen in such a way that the vortex core is not switched. We use 1 to indicate that there is a signal and 0 to indicate that there is no signal (or negligible signal strength) \cite{Jung2012}.\\
\indent First, we considered the case when the IPUA is off in all green disks, thus the signal input is propagated through the neighboring disks, reaching the 2 outputs (disk 1 and disk 7). It is well known that the signal can propagate in disk arrays due to the magnetic interaction between them \cite{Kim:2012,Jung:2011}. The intensity level of the output signals obtained by micromagnetic simulation,  compared to the input signal is shown in (Fig. \ref{salidas1} (a-b)). 

The case where the output signal levels are appreciable, are referred to the logical configuration 111, where the first digit refers to the input signal on disk 4, the second digit refers to the output signal on disk 1, and the third digit refers to the output signal on disk 7. Now, we used the influence of the IPUA in order to cancel the signal on any of the disks used as outputs. As shown in section \ref{twocoupled}, when the directions of the IPUA are perpendicular to the direction to the chain, the magnetic interaction between the disks is reduced.\\
\indent  To get the logical configuration 101, we applied the IPUA in the y direction, only on disk 2 and disk 3. The result is a large decrease in the level of signal on disk 1, while the signal level in disk 7 is high (Fig. \ref{salidas1}(c-d)). The logical configuration 110 is obtained by application of the IPUA in the x direction only on disk 5 and disk 6 (Fig. \ref{salidas1}(e-f)). Now, in order to obtain the logical configuration 100, we applied the IPUA in disks 2, 3, 5 and 6 as follows: on the disk 2 and disk 3 the IPUA is applied in the y direction while, on disk 4 and disk 5 the IPUA is applied in the x direction. With this, we were able to cancel the output signals on disk 1 and disk 7 (Fig. \ref{salidas1}(g-h)). The logical configuration 000 is trivial, since if there is no input signal, there is no output signal.\\
\indent All logical configuration obtained with the Fan-out gate are summarized in Table \ref{tabla1}.

\begin{table}[h!]
\centering
\begin{tabular}{|c|c|c|c|} \hline

Input 	&  Output 1 	& Output 2  &  Logical\\
Disk 4 & Disk 1 & Disk 7 &  configuration\\\hline
0		&	0	& 0 & 000\\\hline

1		&	1	& 1 & 111\\\hline

1	&	0	&   1 & 101\\\hline

1	&	1	& 0 & 110\\\hline

1	&	0	& 0 & 100\\\hline
\end{tabular}
\caption{Logical configuration obtained with the Fan-out gate (Type 1)}\label{tabla1}
\end{table}

\subsubsection{AND, OR gates}
We used the Type 2 array shown in Figure \ref{fanout}. The disk 1 and disk 7 (blue disks) are used as inputs (input 1 and input 2, respectively) and disk 4 as output (red disk). The input signals are implemented by the counterclockwise rotating magnetic field with the same parameters used in subsection \ref{gatelogic}. The value of K$_{\sigma}$ and the directions of the application of the IPUA too are the same used in subsection \ref{gatelogic}.  In an AND gate, the following logical operations can be performed: 0 + 0 = 0, 0 + 1 = 0, 1 + 0 = 0, and 1 + 1 = 1 and in an OR gate, we have: 0 + 0 = 0, 0 + 1 = 1, 1 + 0 = 1, and 1 + 1 = 1 \cite{Omari:2014}.\\

\begin{figure}[h!]
\centering
\includegraphics[width=1\columnwidth]{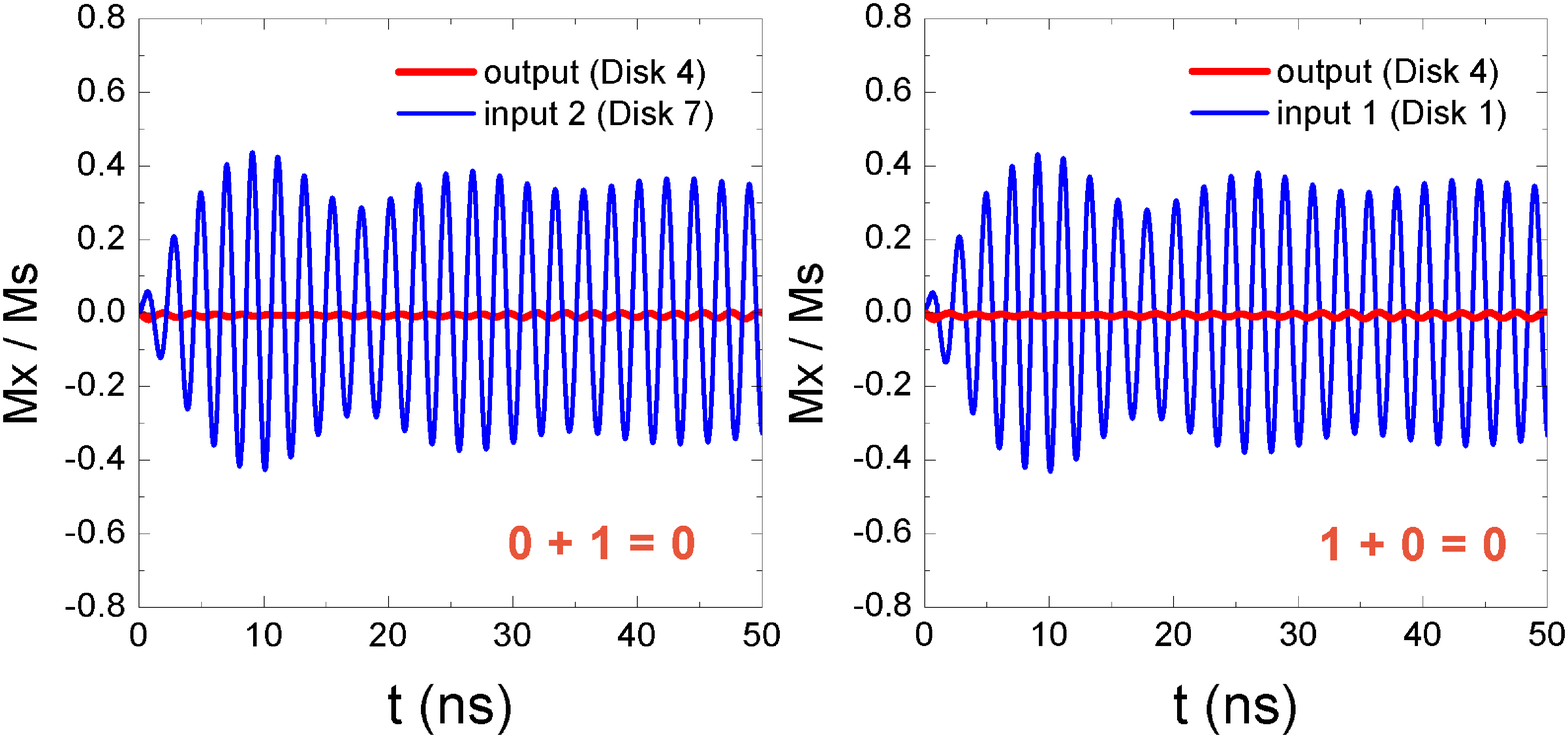}
\includegraphics[width=1\columnwidth]{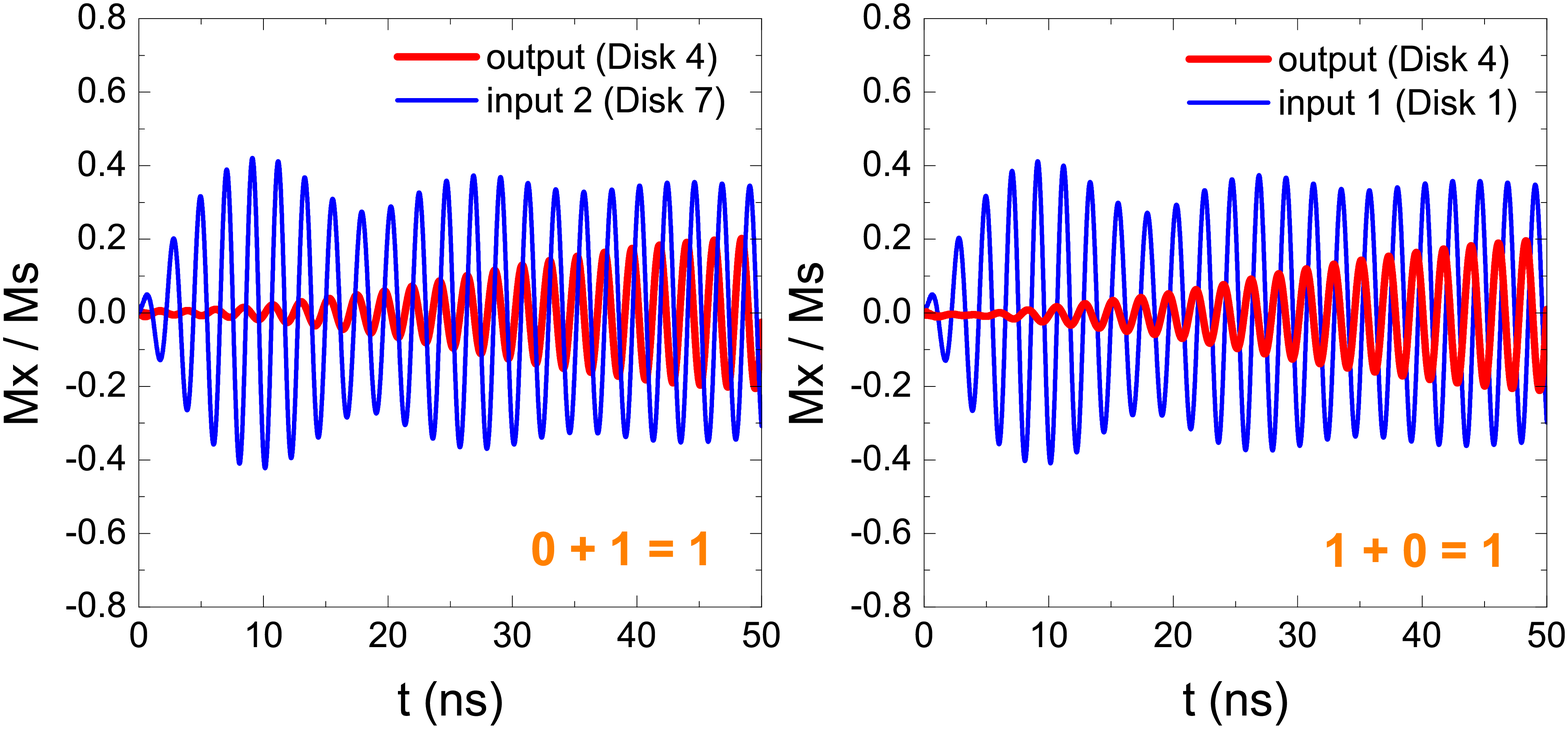}
\caption{Micromagnetic simulation showing a comparison between the intensity levels of the input signals M$_x$/M$_s$ on disk 1 and disk 7 and output signal (disk 4) for an AND and OR gates.}\label{ANDOR}
\end{figure}

\indent  The logical operation 0 + 1  means that we have an input signal only in disk 7, 1 + 0 means there is a input signal only in disk 1 and 1 + 1 is referred when we have input signal in both disk 1 and disk 7. Here again, the operation 0 + 0 is trivial. \\
\indent OR gate: to obtain the logical operation 0 + 1 = 1, it was applied an  input signal only in disk 7. This signal propagates to disk 4. Due to the magnetic interaction, the signal can also propagate to disk 1 (input disk), but to avoid this propagation, we have applied the IPUA in disk 2 and disk 3. Similarly, to obtain the logical operation 1 + 0 = 1, an input signal was supplied to disk 1 and the IPUA was applied in disk 5 and disk 6. These results can be seen in Fig. \ref{ANDOR}.  \\
\indent AND gate: To obtain the logical operation 0 + 1 = 0, it was applied an input signal only in disk 7. In order to avoid the propagation of the signal, the IPUA was applied in disk 5 and disk 6. In a similar way, to obtain the logical operation 1 + 0 = 0, an input signal was applied only in disk 1, and IPUA in disk 2 and disk 3. These results are shown in Fig. \ref{ANDOR}.\\
\indent The logical operation 1 + 1 = 1 is the same for both AND and OR gates; this means having an input signal in disk 1 and disk 7. In this case, it is not necessary to apply the IPUA, since we want to have an output signal in disk 4. This operation can be considered trivial and  is not shown here.\\
\indent The advantage of using the IPUA, is that using it, we can cancel the propagation along a chosen direction. This would be impossible to do with some other external agent acting on the nanodisks and preserving the vortex configuration.

\section{Conclusions}
In this work, the influence of the direction of application of the IPUA in a system of two coupled identical nanodisks with magnetic vortex configuration was first studied using micromagnetic simulation. We demonstrated that it is possible to obtain shorter energy transfer times when the IPUA is along the x axis, while if the IPUA is along the y axis, the magnetic interaction between the disks is canceled, therefore the energy transfer is not efficient. Also, using a simple dipolar model, we obtained the coupling integrals  depending on K$_\sigma$, and we explained  why $\tau$ increases or decreases, depending on the direction of application of the IPUA.\\
\indent We have created Fan-out, AND and OR gates in an array of nanodisks using the influence of the IPUA to control the propagation of the signal in a desired direction.  All these results show the great efficiency of using IPUA to control the magnetic interaction and the direction of  signal propagation between the disks without the need of changing the separation distance between them.\\
\indent These results show the large potential of using magnetic vortices and controlled in-plane anisotropy for the design of logic gates and other devices.
\section*{ACKNOWLEDGMENTS}
The authors would like to thank the support of the Brazilian agencies CNPq and FAPERJ.

\section*{References}

\bibliography{referencias}

\end{document}